\documentclass[10pt,a4paper,fleqn]{article}

\usepackage[
    top=20mm,
    bottom=20mm,
    left=18mm,
    right=18mm
]{geometry}

\usepackage{graphicx}
\usepackage{svg}
\usepackage{multicol}
\usepackage{titlesec}
\usepackage{fancyhdr}
\usepackage{setspace}
\usepackage{array}
\usepackage{booktabs}
\usepackage{tikz}
\usepackage{tcolorbox}
\usepackage[
    font=scriptsize,
    labelfont=bf,
    justification=justified,
    singlelinecheck=false
]{caption}
\usepackage{fontspec}
\usepackage{amsmath}
\usepackage{amsfonts}
\usepackage{siunitx}
\usepackage{url}

\usepackage{titlesec}
\titleformat{\section}
    {\normalsize\bfseries}
    {\thesection}
    {1em}
    {}
\titleformat{\subsection}
    {\small\bfseries}
    {\thesubsection}
    {1em}
    {}
\titleformat{\paragraph}[runin]
    {\footnotesize\bfseries}
    {\theparagraph}
    {1em}
    {}

\usepackage{tabularx}
\newcolumntype{R}{>{\raggedleft\arraybackslash}X}

\usepackage{lipsum}

\definecolor{tether-gray}{HTML}{F2F1EF}

\let\oldfootnote\footnote
\renewcommand{\footnote}[1]{\oldfootnote{\scriptsize #1}}

\newtcolorbox[auto counter]{custombox}[2][]{
    title={Box~\thetcbcounter: #2},
    label={#1},
    fonttitle=\bfseries,
    coltitle=black,
    before title=\vspace{15pt},
    colback=tether-gray,
    colframe=tether-gray,
    boxrule=0pt,
    arc=5pt,
    left=15pt,
    right=15pt,
    bottom=15pt,
    width=\textwidth,    
}

\newcommand{\tablebox}[1]{
    \tcbox[
        colback=tether-gray,
        colframe=tether-gray,
        boxrule=0pt,
        left=15pt,
        right=15pt,
        top=15pt,
        bottom=15pt
    ]{#1}
}

\newcommand{\model}{BrainWhisperer}

\begin{document}


\begin{flushleft}
\includegraphics[height=20pt]{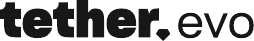}
\end{flushleft}

\vspace{10pt}


\begin{tcolorbox}[
    colback=tether-gray,
    colframe=tether-gray,
    boxrule=0pt,
    arc=5pt,
    left=15pt,
    right=15pt,
    top=15pt,
    bottom=15pt,
    width=\textwidth
]


{\huge\bfseries\begin{flushleft}
\model: Leveraging Large-Scale ASR Models for Neural Speech Decoding
\end{flushleft}}


{\small\bfseries
Tommaso Boccato\textsuperscript{1},
Michal Olak\textsuperscript{1},
Matteo Ferrante\textsuperscript{1}
}

\vspace{2.5pt}

{\small
\textsuperscript{1}Tether Evo
}

\vspace{15pt}


{\footnotesize
Decoding continuous speech from intracortical recordings is a central challenge for brain-computer interfaces (BCIs), with transformative potential for individuals with conditions that impair their ability to speak. While recent microelectrode array (MEA) decoders achieve impressive accuracy, their performance is fundamentally limited by the small size of existing datasets, they remain brittle to session-to-session variability, and their ability to generalize across participants remains unexplored. We introduce \model{}, a neural speech decoder that integrates high-resolution MEA recordings with a large pretrained automatic speech recognition (ASR) model. Building on interpretability findings showing that Whisper's encoder learns phoneme-selective representations with localized attention, we train a customized version of Whisper, modified to process neural features, using a hybrid objective that combines CTC loss on phonemes--predicted from the third encoder layer--and cross-entropy loss on word tokens. We introduce domain-informed modifications including windowed self-attention to capture articulatory continuity, hierarchical month/day-specific low-rank projections to address non-stationarity, and subject-specific embedders enabling cross-subject training. Evaluated on a publicly available MEA dataset (Card et al.), \model{} matches or outperforms prior state-of-the-art decoders. Critically, cross-dataset training improves performance even on individual datasets without fine-tuning, demonstrating unprecedented generalization. The model supports dual decoding paths: a high-accuracy phoneme-based path with external language model rescoring, and a fast direct text generation path enabling sub-100ms inference with minimal hardware requirements. By bridging large-scale ASR pretraining with neural data, our work demonstrates a scalable pathway toward foundation models for speech BCIs, addressing key barriers of data scarcity, computational cost, and cross-participant generalization while raising important considerations about neural privacy and user agency.
}

\end{tcolorbox}

\vspace{10pt}


\footnotesize

\begin{figure}[h]
    \hspace{15pt}
    \includegraphics[width=\dimexpr\textwidth-30pt-5pt]{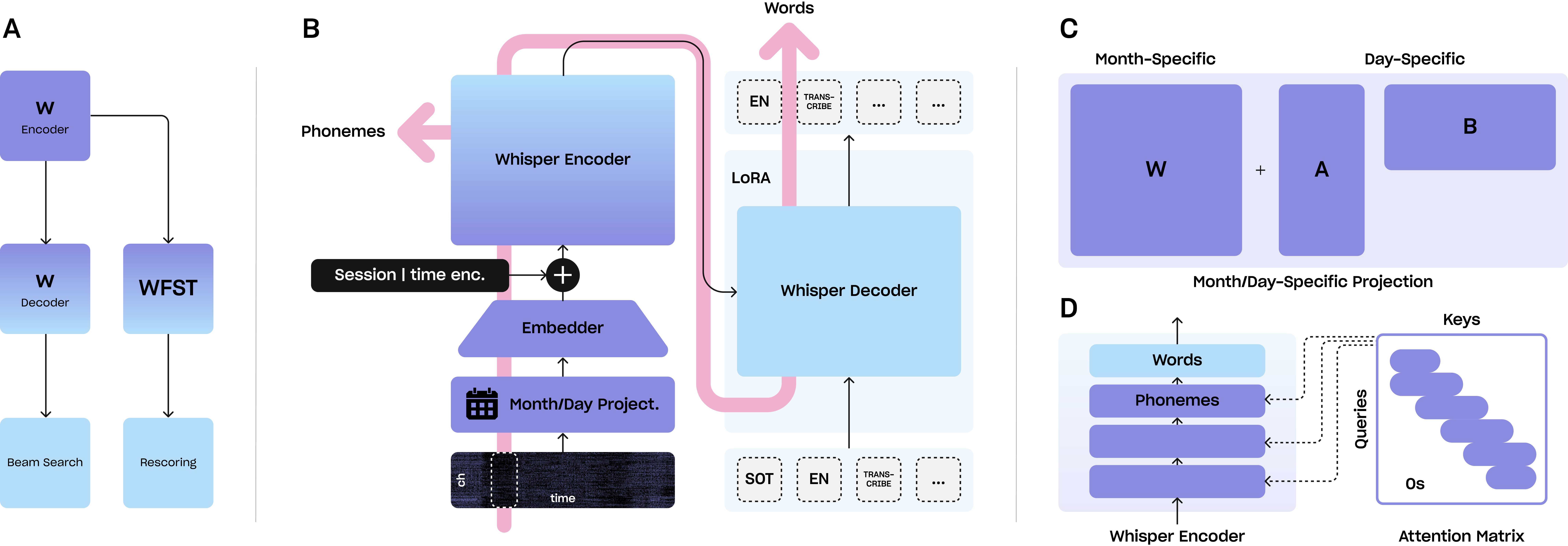}
    \caption{\textbf{A.} Our decoding pipelines: the low-compute end-to-end branch uses an encoder–decoder architecture with beam search, while the high-compute branch feeds phoneme logits into a weighted finite-state transducer to generate candidates, which are then rescored. \textbf{B.} \model{}'s architecture. Neural features are first mapped into a shared embedding space using a convolutional front-end (embedder) and then processed by a stack of transformer encoder layers. A phoneme head predicts phoneme sequences using a CTC loss, while the remaining encoder layers feed a transformer decoder trained with cross-entropy on text tokens. \textbf{C.} The proposed month-specific and day-specific low-rank projections. \textbf{D.} Our windowed attention mechanism. Windowing is applied to the phoneme-selective encoder layers.}
    \label{fig:whisper_model}
\end{figure}

\section*{Background and Motivation}

Neural speech decoding aims to restore communication for individuals with severe paralysis by translating cortical activity into text or speech. While recent MEA-based decoders achieve word error rates (WERs) below 10\% \cite{willett2023high,Card2024,zhang2026decoding}, they face critical limitations: extreme data requirements ($\sim$10$^4$ sentences per participant), rapid performance degradation across recording sessions due to non-stationarity, and inability to generalize across individuals. Current state-of-the-art systems use recurrent networks with CTC phoneme prediction followed by heavy n-gram language model rescoring, requiring hundreds of gigabytes of RAM and making deployment on privacy-preserving local hardware impractical. We hypothesize that leveraging pretrained ASR models can address these challenges by providing robust phonetic representations and linguistic priors learned from massive speech corpora.

\section*{Methods}

We adapted Whisper \cite{pmlr-v202-radford23a}, an ASR model pretrained on 680k hours of speech, to decode neural activity from two BrainGate2 participants \cite{willett2023high,Card2024,kunz2025innerspeech} implanted with 256-channel Utah arrays in speech motor cortex (Figure \ref{fig:whisper_model}). Neural features (spike counts and spike-band power at 50Hz) undergo hierarchical ``normalization'': month-level projections capture slow population shifts, while low-rank day-specific corrections address finer variability and prevent overfitting. Transformed features pass through convolutional layers producing token sequences fed to a modified Whisper encoder. Critically, motivated by findings that early Whisper layers learn phoneme-selective features with localized receptive fields, we replace global attention in the first three encoder layers with non-causal windowed attention, encoding the inductive bias that articulatory dynamics are locally constrained while reducing computational complexity.\footnote{The observed reduction may vary depending on the specific attention implementation.} A phoneme prediction head attached at layer 5 (the encoder has a total of 6 layers) is trained with CTC loss on phoneme sequences, while the remaining encoder and decoder layers (we use LoRA \cite{hu2021loralowrankadaptationlarge} for fine-tuning the Whisper's decoder) are trained with cross-entropy loss on text tokens. This dual objective ($\mathcal{L} = \mathcal{L}_{CTC} + \mathcal{L}_{CE}$) encourages representations that are simultaneously phoneme-discriminative and linguistically coherent. We found it beneficial to split the training steps into two separate phases, one for the computation of each loss term; time masking from \cite{feghhi2025timemasked} was applied during the CTC loss term computation and helped prevent overfitting. The model supports two decoding paths: (1) phoneme-based decoding with external weighted finite-state transducer (WFST) and 5-gram rescoring for maximum accuracy, and (2) direct greedy text generation from Whisper's decoder for low-latency inference. We refer the reader to the Appendix for a detailed description of our methodology.

\section*{Results and Discussion} 

We trained models on \textit{``Card''} data \cite{Card2024} alone and jointly on \textit{``Card+Willett+Kunz''} data \cite{Card2024,willett2023high,kunz2025innerspeech} ($\sim$18,000 training trials total), evaluating on held-out test and \textit{Brain-to-Text '25}\footnote{\url{https://www.kaggle.com/competitions/brain-to-text-25}} competition splits (Table \ref{tab:accuracy}). Our base model surpasses the performance of the RNN baseline \cite{willett2023high,Card2024,zhang2026decoding} in both phoneme-based decoding and end-to-end (E2E) generation, with an even larger margin in the latter case. The model nearly matches the WER achieved by the corresponding BIT \cite{zhang2026decoding} in phoneme-based decoding and outperforms it when considering the E2E decoding path. With cross-dataset training, we remain approximately 1 percentage point behind the corresponding BIT, yet we clearly outperform it in terms of E2E generation WER, achieving an impressive 8.7\% (see Table \ref{tab:examples} for some decoding examples), which, to the best of our knowledge, represents the best E2E decoding performance ever reached on the benchmark used. This result is particularly significant because it not only demonstrates the effectiveness of Whisper’s integrated decoder in providing robust language modeling but also, most importantly, narrows the gap between phoneme-based and E2E decoding. For the first time, it drives performance below the psychological E2E WER threshold of 10\%, a value under which models can be considered for real-world deployment. Notably, cross-dataset training improved performance on the original Card dataset \textbf{without fine-tuning}. The computational advantages are substantial: E2E decoding requires less than 2GB of VRAM with approximately 50ms inference time, compared to approximately 300GB of RAM and 750ms for n-gram rescoring, enabling practical local deployment. Ablation studies confirmed that windowed attention, dual loss objectives, and low-rank projections each contribute meaningfully to overall performance.

\begin{table}[ht]
\caption{Accuracy in terms of phoneme error rate (PER) and word error rate (WER) for the evaluated models. Results are shown for both the test and competition splits of the Card dataset. The \textit{5-gram} subscript indicates that extensive post-processing with a 5-gram model and rescoring was applied.}
\label{tab:accuracy}
\footnotesize{
\tablebox{
\begin{tabularx}{\dimexpr\textwidth-30pt-5pt}{l *{3}{R} *{2}{R}}
\toprule
 & \multicolumn{3}{c}{\textbf{Test}} & \multicolumn{2}{c}{\textbf{Competition}} \\ 
\cmidrule(lr){2-4}\cmidrule(lr){5-6}
\textbf{Model} & \textbf{PER} & \textbf{WER\textsubscript{5-gram}} & \textbf{WER\textsubscript{E2E}} & \textbf{WER\textsubscript{5-gram}} & \textbf{WER\textsubscript{E2E}} \\ 
\midrule
RNN & 9.6\% & n/a & n/a & 6.7\% & 16.0\% \\
BIT & 8.9\% & n/a & n/a & 5.2\% & 12.2\% \\
BIT (cross) & \textbf{7.1}\% & n/a & n/a & \textbf{4.1}\% & 11.1\% \\
Ours & 8.6\% & \textbf{5.4}\% & 9.8\% & 5.9\% & 10.2\% \\
Ours (cross) & 9.5\% & \textbf{5.4}\% & \textbf{8.9}\% & 5.2\% & \textbf{8.7}\% \\
\bottomrule
\end{tabularx}
}
}
\end{table}

\begin{table}[ht]
\caption{Examples of decoded transcriptions. Each example reports the ground truth transcription, the one predicted by the model and the corresponding WER. We selected examples associated with the 10\textsuperscript{th}, 20\textsuperscript{th}, \dots, 100\textsuperscript{th} percentiles. Errors are highlighted in red.}
\label{tab:examples}
\footnotesize{
\includegraphics[width=.65\textwidth]{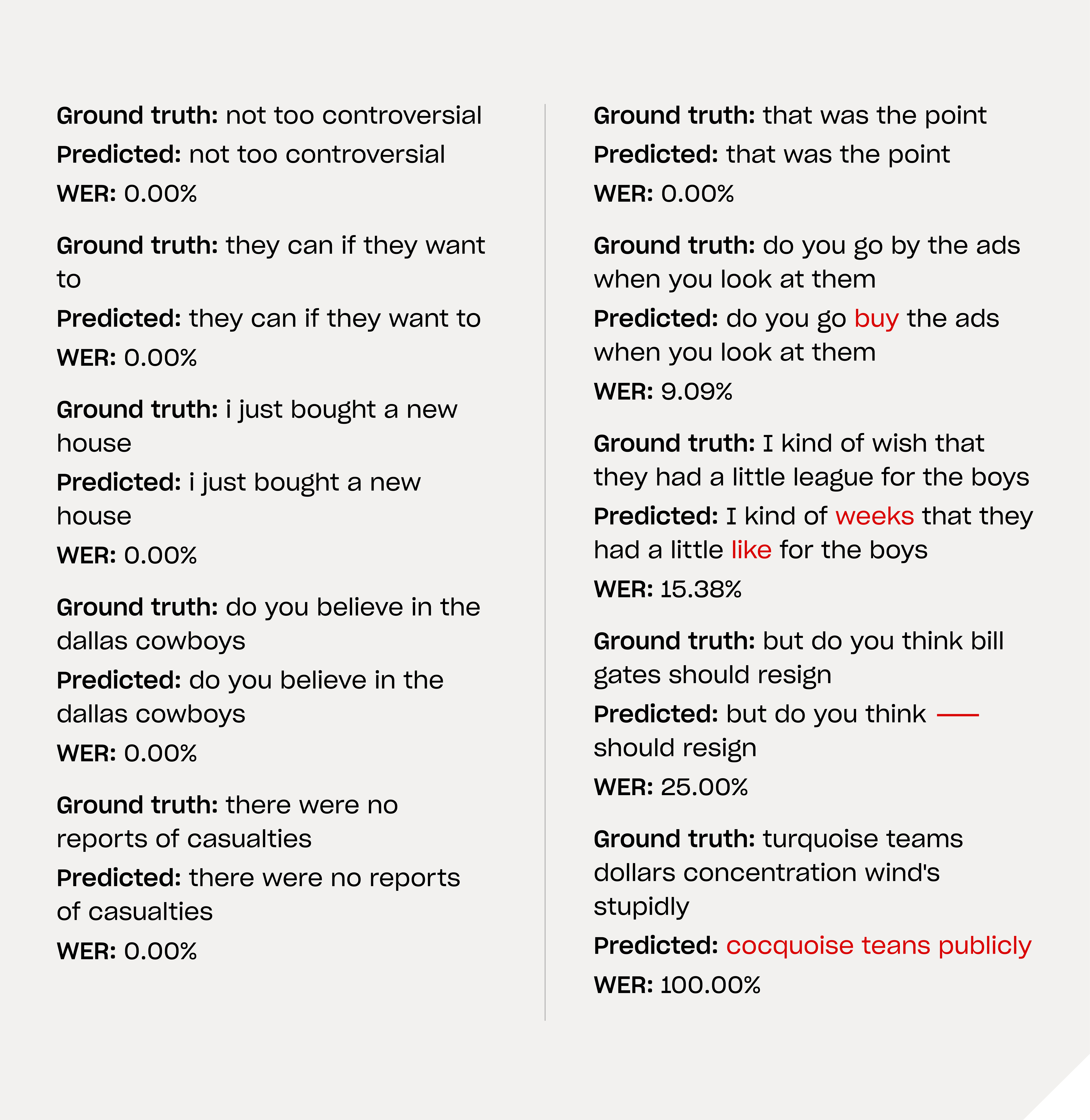}
}
\end{table}

\section*{Ethical Considerations} 

Neural data represents extremely private information. We emphasize that successful decoding requires user cooperation and can be disrupted by deliberate cognitive strategies (mental counting, inner monologue), providing users with volitional control. The lightweight computational footprint of our decoder enables local inference without cloud transmission, addressing critical privacy concerns. Future work should explore mechanisms like ``mental passwords'' and adaptive confidence thresholds to ensure systems remain under user agency.

\section*{Brain-to-Text '25 Challenge}

An extension of our high-compute pipeline secured 4\textsuperscript{th} place out of 466 teams in the Brain-to-Text '25 Kaggle challenge, decoding 256-channel neural data without alignment. Meanwhile, our jointly designed low-compute branch—optimized for local inference, latency, and privacy—achieves state-of-the-art end-to-end decoding performance.

\section*{Code Availability}

A link to the project's repository will be provided in a future version of the manuscript.


\bibliographystyle{ieeetr}
\bibliography{bib}


\appendix

\section{Model Architecture}

Let $X \in \mathbb{R}^{C \times T}$ denote the multivariate time-series to be decoded. $X$ consists of $T$ temporal frames $\boldsymbol{x}_t$, each containing $C$ neural features computed from signals recorded across multiple channels, as described in \cite{willett2023high,Card2024}.

\subsection{Neural Embedder}

The first component of our approach renders the input data digestible to Whisper's encoder by embedding it into a sequence of tokens. This process is performed by an embedder composed of two distinct modules: a projector designed to mitigate data non-stationarity, and a sequence of two convolutional layers responsible for producing the actual tokens.

\paragraph{Month/Day-Specific Projections}

Our first contribution lies in what we call month/day-specific projections. In prior work \cite{willett2023high,Card2024}, each temporal frame is mapped into a common space across multiple recording sessions through linear transformations learned during training. While this approach is effective when a large number of trials per session is available, it yields diminished gains when trials per session are limited. To circumvent this problem, we developed a new type of linear transformation (followed by a non-linearity $\sigma$), given by the equation:
\begin{equation}
    \tilde{\boldsymbol{x}}_t = \sigma\Big(\big[W + (A \cdot B)\big] \boldsymbol{x}_t + \boldsymbol{b}_{mth} + \boldsymbol{b}_{day}\Big)
\end{equation}
The projection matrix is thus the sum of two distinct terms: a full-rank matrix $W$ common to all sessions belonging to the same month (the entire data acquisition was divided into 30-day temporal segments), and a low-rank matrix obtained as the product of $A \in \mathbb{R}^{C \times R}$ and $B \in \mathbb{R}^{R \times C}$ with $R \ll C$, specific to the considered recording day. The biases $\boldsymbol{b}_{mth} \in \mathbb{R}^C$ and $\boldsymbol{b}_{day} \in \mathbb{R}^C$ follow the same scheme. The use of these projections was coupled with a batch sampler capable of guaranteeing the presence of 16 samples from each session within a training batch.

\paragraph{Convolutional Token Generation}

\sloppypar{
The embedder proceeds with the two aforementioned convolutions, producing a sequence of embeddings $E = [\dots, \boldsymbol{e}_\tau, \dots] \in \mathbb{R}^{D \times T'}$, where $T'$ depends on the stride of the second convolution and $D$ is the hidden dimension used by the remaining architecture.
}

\paragraph{Positional Encodings}

At this point, the model receives information about the day (session) and temporal frame of origin for each token in the form of sinusoidal positional encodings. Specifically, the new embeddings with positional encodings are obtained as:
\begin{equation}
    \tilde{E} = \bigg[\dots, \boldsymbol{e}_\tau + \Big[{\boldsymbol{e}_s^{pos}}^T\ |\ {\boldsymbol{e}_\tau^{pos}}^T\Big]^T, \dots\bigg]
\end{equation}
where $\boldsymbol{e}_s^{pos}, \boldsymbol{e}_\tau^{pos} \in \mathbb{R}^\frac{D}{2}$, $s$ represents the session from which the features in $X$ originate, and $\tau$ denotes time as defined in the context of Whisper's encoder.

\subsection{Modified Whisper Encoder}

The embeddings now serve as input to a Whisper encoder, identical to the original except for two important modifications.

\paragraph{Phoneme Prediction Head}

First, building on interpretability studies \cite{Reid2023InterpretingWhisper} showing that Whisper's encoder transitions from encoding phoneme representations to word representations as layers increase, we attached a phoneme prediction head at intermediate layer $L' < L$, where $L$ is the number of encoder layers. This prediction head is simply a linear layer operating on the output of layer $L'$ where embeddings have been concatenated in pairs (thus producing one prediction every 80ms).

\paragraph{Masked Attention}

Furthermore, all layers up to layer $L'$ employ windowed (non-causal) attention. In other words, each position in a layer can attend to a limited temporal window of tokens produced by the previous layer, centered on the current token:
\begin{equation}
    \tilde{A}^{att} = M \odot A^{att},\quad M_{ij} =
    \begin{cases}
        1\quad \text{if}\ |i - j| \le w\\
        0\quad \text{otherwise}
    \end{cases}
\end{equation}
Here, $A^{att}$ is the generic attention matrix present in all layers up to the ${L'}$-\textit{th}, while $M$ is the mask. The rationale is to help the model focus on the information strictly necessary to solve the phoneme decoding sub-task, based on our hypothesis that the signals guiding articulation, as a first approximation, do not need to encode long context for successful word production.

\subsection{Whisper Decoder}

\sloppypar{
As decoder, we retained the original pre-trained Whisper decoder, fine-tuned during training with LoRA \cite{hu2021loralowrankadaptationlarge}. This choice allowed us to overcome the impossibility of learning language statistics from the limited available data and to bypass the conditional independence assumption required by CTC-based approaches. The decoder receives as input Whisper's beginning of sequence (BOS) tokens (\texttt{<|startoftranscript|><|en|><|transcribe|><|notimestamps|>}) and all tokens generated up to the current moment, cross-attends to the tokens produced by the encoder, and finally generates the next token. Having implemented the model following HuggingFace's generation interfaces, it is compatible out-of-the-box with the library's various generation strategies.
}

\section{Training Framework}

Let's now consider the training framework.

\subsection{Multi-Task Objective}

As already mentioned, a multi-task training framework was designed for the model in which two different objectives are optimized simultaneously: a connectionist temporal classification (CTC) loss for phoneme prediction and a cross-entropy (CE) loss for next-token prediction of words:
\begin{equation}
    \mathcal{L} = \mathcal{L}_{CTC} + \mathcal{L}_{CE}
\end{equation}

\subsection{Two-Phase Training Step}

To prevent overfitting, we implemented the augmentations described in \cite{Card2024} and employed the masking proposed in \cite{feghhi2025timemasked}. In preliminary experiments, we observed that using masking concurrently with optimization of the CE loss led to instability in the loss itself, while disabling masking resulted in generally sub-optimal performance. We therefore resolved this issue through an original reformulation of the multi-task framework in which the training step was divided into two phases. In the first phase, the model is executed with masking active up to the phoneme prediction head, enabling calculation of the information necessary for the CTC loss while preventing overfitting. In the second phase, the model is re-executed with masking disabled in order to compute the second term of the global loss, the CE. In this way, at each training step it is possible to update all learnable weights of the model using samples from the same batch while exploiting the optimal setup for each objective.

\section{Cross-Dataset Configuration}

Part of the experiments were conducted in cross-dataset mode, meaning training and testing were performed on different datasets, and thus different tasks, different subjects, and different numbers of electrodes/channels across subjects. In these cases, this variability was managed through the month/day-specific projections, using different transformations for each unique session (and thus subjects). This approach allows for straightforward mapping of different data into the same space, before proceeding with the remaining shared architecture. We note that in cross-dataset mode, the batch sampler must also be updated to handle the presence of multiple datasets within the same batch. Specifically, we maintained 16 samples per session and sampled sessions uniformly within the same dataset, allocating on average for each dataset a number of sessions proportional to the datasets' respective sizes.

\end{document}